\documentstyle[12pt]{article}

\newcommand{\eq}{\begin{equation}}
\newcommand{\eqa}{\begin{eqnarray}}
\newcommand{\en}{\end{equation}}
\newcommand{\ena}{\end{eqnarray}}
\newcommand{\enn}{\nonumber \end{equation}}

\def\sk{\vskip .4cm}
\def\noi{\noindent}

\def\epsi{\varepsilon}

\def\de{\delta}

\def\part{\partial}

\def\R#1#2{ R^{#1}_{~~~#2} }
\def\PA#1#2{ P^{#1}_{A~~#2} }

\def\Rinv#1#2{ (R^{-1})^{#1}_{~~~#2} }

\def\Rb{{\bf R}}
\def\Rbo{{\bf R}}

\def\Rh{{\hat R}}

\def\Rhat#1#2{ \Rh^{#1}_{~~~#2} }

\def\Rhatinv#1#2{ (\Rh^{-1})^{#1}_{~~~#2} }

\def\T#1#2{ T^{#1}_{~~#2} }

\def\rminus{r^{-1}}

\def\D{\Delta}

\def\IGLqrN{IGL_{q,r}(N)}

\def\GLqrNo{GL_{q,r}(N+1)}
\def\SOqrNt{SO_{q,r}(N+2)}
\def\SpqrNt{Sp_{q,r}(N+2)}

\def\ISOqrN{ISO_{q,r}(N)}
\def\ISpqrN{ISp_{q,r}(N)}
\def\ISOqroN{ISO_{q,r=1}(N)}
\def\ISpqroN{ISp_{q,r=1}(N)}

\def\ISqrN{IS_{q,r}(N)}

\def\SqrNt{S_{q,r}(N+2)}

\def\Tc{{\cal T}}

\def\Dtwo{\Delta_{N+2}}
\def\epsitwo{\epsi_{N+2}}
\def\kappatwo{\kappa_{N+2}}

\def\n2{{{N+1} \over 2}}
\def\ap{a^{\prime}}
\def\bp{b^{\prime}}
\def\cp{c^{\prime}}
\def\dpr{d^{\prime}}
\def\Dc{{\cal D}}

\def\Ntwo{{N\over 2}}

\def\square{{\,\lower0.9pt\vbox{\hrule \hbox{\vrule height 0.2 cm
\hskip 0.2 cm \vrule height 0.2 cm}\hrule}\,}}

\begin{document}

\begin{titlepage}
\rightline{DFTT-44/94}
\rightline{November 1994}
\vskip 2em
\begin{center}{\bf   R-MATRIX FORMULATION OF THE
QUANTUM INHOMOGENEOUS GROUPS   $ISO_{q,r}(N)$
AND $ISp_{q,r}(N)$.}
\\[3em]
Paolo Aschieri \\[1em]
{\sl Scuola Normale Superiore\\
Piazza dei Cavalieri 7,  56100 Pisa \\and\\Istituto Nazionale
di Fisica Nucleare,\\ Sezione di Pisa, Italy}\\[3em]
Leonardo Castellani\\[1em]
{\sl II Facolt\`a di Scienze M.F.N. di Torino, sede di
Alessandria}\\[.5 em]
{\sl Dipartimento di Fisica Teorica\\
and\\Istituto Nazionale di
Fisica Nucleare\\
Via P. Giuria 1, 10125 Torino, Italy.}  \\[3em]
\end{center}

\begin{abstract}
The quantum commutations $RTT=TTR$ and
the orthogonal (symplectic) conditions for
the inhomogeneous multiparametric $q$-groups
of the $B_n,C_n,D_n$ type are found in terms of
the $R$-matrix of  $B_{n+1},C_{n+1},D_{n+1}$.
A consistent Hopf structure on these
inhomogeneous $q$-groups is constructed by means of
a projection from $B_{n+1},C_{n+1},D_{n+1}$.
Real forms are discussed: in particular we obtain the $q$-groups
$ISO_{q,r}(n+1,n-1)$, including the
quantum Poincar\'e group.
\end{abstract}

\vskip 2cm

\noi DFTT-44/94

\noi  November1994~~~~~~~~~~~~~~~~~~~~~~~~~~~~~~~~~~~~~~~
{}~~~~~~~~~~~~~~~~~~~~~~~~~~~~~~~~~hep-th/9411039
\vskip .2cm
\noi \hrule
\vskip.2cm
\hbox{\vbox{\hbox{{\small{\it e-mail: }}}\hbox{}}
 \vbox{\hbox{{\small decnet=31890::aschieri,
31890::castellani}}
\hbox{{\small internet = aschieri@ux2sns.sns.it,
castellani@to.infn.it
}}}}

\end{titlepage}
\newpage
\setcounter{page}{1}

%\sect{Introduction}

Inhomogeneous groups play an important role
in many physical situations, for  instance when translations enter
the game.
One fundamental example is Einstein-Cartan gravity, whose algebraic
basis
is the Poincar\'e group. After the discovery of $q$-deformed
simple Lie groups \cite{qgroups1,FRT}, it was natural to construct
the corresponding $q$-deformed gauge theories \cite{qgauging,Cas1}.
A similar program can be applied to inhomogeneous $q$-groups,
and indeed in ref. \cite{Cas2} a $q$-deformation of Poincar\'e
gravity
was found, one of  the main motivations being the possibility of
$q$-regularizing gravity. The gauge program relies on the bicovariant
calculus on $q$-groups: for a review see for
ex. \cite{Aschieri} and references quoted therein.
\sk
We present in this Letter the $R$-matrix formulation
of multiparametric inhomogeneous $q$-groups, whose homogeneous part
are the $B_n,C_n,D_n$ $q$-groups. This extends
to the orthogonal and symplectic case the treatment of ref.
\cite{Cas3,Aschieri1},
where the multiparametric (uniparametric in \cite{Cas3})
 $q$-groups $\IGLqrN$ and their associated differential
calculi
were constructed via a projection from $\GLqrNo$. Some of the
references
on the quantum inhomogeneous groups
are collected in \cite{inhom}.
\sk
The method used in \cite{Cas3,Aschieri1} to obtain $\IGLqrN$,
and in this Letter to obtain $ISO_{q,r}(N)$
and $ISp_{q,r}(N)$, is based on a consistent
projection from the corresponding quantum groups
of higher rank $A_{n+1}, B_{n+1}, C_{n+1}, D_{n+1}$.
By consistent we mean that it is compatible (or ``commutes")
with the Hopf structure of the  $q$-groups,
as we will see in the sequel.  This method was in fact already
exploited in ref. \cite{Cas2} to obtain bicovariant
 differential calculi on
$\ISOqroN$ and $\ISpqroN$.
\sk
We give here the explicit structures of  $ISO_{q,r}(N)$
and $ISp_{q,r}(N)$, and show that they are Hopf algebrae
by proving that they can be obtained as quotients
of $\SOqrNt$ and $\SpqrNt$ with respect
to a suitable Hopf ideal. The projection
from $\SOqrNt$  [$\SpqrNt$] to the quotient is introduced
and found to be an Hopf algebra epimorphism.
The (bicovariant)
differential calculi on the multiparametric
$\ISOqrN$ [$\ISpqrN$], found by means of this
projection method, will be
presented in a separate publication.
\sk
We begin by recalling some basic facts about the $B_n, C_n, D_n$
 multiparametric
quantum groups. They are freely generated by the noncommuting
matrix elements $\T{a}{b}$ (fundamental representation) and
the identity $I$. The noncommutativity is controlled by the $R$
matrix:
\eq
\R{ab}{ef} \T{e}{c} \T{f}{d} = \T{b}{f} \T{a}{e} \R{ef}{cd}
\label{RTT}
\en
which satisfies the quantum Yang-Baxter equation
\eq
\R{a_1b_1}{a_2b_2} \R{a_2c_1}{a_3c_2} \R{b_2c_2}{b_3c_3}=
\R{b_1c_1}{b_2c_2} \R{a_1c_2}{a_2c_3} \R{a_2b_2}{a_3b_3}, \label{QYB}
\en
a sufficient condition for the consistency of the
``$RTT$" relations (\ref{RTT}).  The $R$-matrix components
$\R{ab}{cd}$
depend
continuously on a (in general complex)
set of  parameters $q_{ab},r$.  For $q_{ab}=q, r=q$ we
recover the uniparametric $q$-groups of ref. \cite{FRT}. Then
$q_{ab} \rightarrow 1, r \rightarrow 1$ is the classical limit for
which
$\R{ab}{cd} \rightarrow \de^a_c \de^b_d$ : the
matrix entries $\T{a}{b}$ commute and
become the usual entries of the fundamental representation. The
multiparametric $R$ matrices for the $A,B,C,D$ series can be
found in \cite{Schirrmacher}  (other ref.s on multiparametric
$q$-groups are given in \cite{multiparam1,multiparam2}). For
the $B,C,D$ case they read:
\eq
\begin{array}{ll}
\R{ab}{cd}=&\delta^a_c \delta^b_d [{r\over q_{ab}} +
(r-1) \delta^{ab}+(\rminus - 1)
\de^{a\bp}] (1-\de^{a n_2})
+\de^a_{n_2} \de^b_{n_2} \de^{n_2}_c \de^{n_2}_d  \\
&+(r-r^{-1})
[\theta^{ab} \delta^b_c \de^a_d - \epsilon_a\epsilon_c
\theta^{ac} r^{\rho_a - \rho_c}
\de^{\ap b} \de_{\cp d}]
\end{array}
\label{Rmp}
\en
\noi where $\theta^{ab}=1$ for $a> b$
and $\theta^{ab}=0$ for $ a \le b$; we define
 $n_2 \equiv \n2$ and primed indices as $\ap \equiv N+1-a$. The
indices
run on $N$ values ($N$=dimension of
the fundamental representation $\T{a}{b}$),
with $N=2n+1$ for $B_n  [SO(2n+1)]$,  $N=2n$
for $C_n [Sp(2n)]$, $D_n [SO(2n)]$.
The terms with the index $n_2$ are present
only for the $B_n$ series. The $\epsilon_a$ and
$\rho_a$ vectors are given by:
\eq
\epsilon_a=
\left\{ \begin{array}{ll} +1 & \mbox{for $B_n$, $D_n$} ,\\
                            +1 & \mbox{for $C_n$ and $a \le n$},\\
                              -1  & \mbox{for $C_n$ and $a > n$}.
           \end{array}
                                    \right.
\en
\eq
(\rho_1,...\rho_N)=\left\{ \begin{array}{ll}
         (\Ntwo -1, \Ntwo -2,...,
{1\over 2},0,-{1\over 2},...,-\Ntwo+1)
                   & \mbox{for $B_n$} \\
           (\Ntwo,\Ntwo -1,...1,-1,...,-\Ntwo) & \mbox{for $C_n$} \\
           (\Ntwo -1,\Ntwo -2,...,1,0,0,-1,...,-\Ntwo+1) & \mbox{for
$D_n$}
                                             \end{array}
                                    \right.
\en
Moreover the following relations reduce the number of independent
$q_{ab}$ parameters \cite{Schirrmacher}:
\eq
q_{aa}=r,~~q_{ba}={r^2 \over q_{ab}}; \label{qab1}
\en
\eq
q_{ab}={r^2 \over q_{a\bp}}={r^2 \over q_{\ap b}}=q_{\ap\bp}
 \label{qab2}
\en
\noi where (\ref{qab2}) also implies $q_{a\ap}=r$. Therefore
 the $q_{ab}$ with $a < b \le {N\over 2}$ give all the $q$'s.
\sk
It is useful to list the nonzero complex components
of the $R$ matrix (no sum on repeated indices):
\eqa
& &\R{aa}{aa}=r , ~~~~~~~~~~~~~~\mbox{\footnotesize
 $a \not= {n_2}$ } \cr
& &\R{a\ap}{a\ap}=r^{-1} ,  ~~~~~~~~~~\mbox{\footnotesize
 $a \not= {n_2}$ } \cr
& &\R{{n_2}{n_2}}{{n_2}{n_2}}= 1\cr
& &\R{ab}{ab}={r \over q_{ab}} ,~~~~~~~~~~~~\mbox{\footnotesize
 $a \not= b$,  $\ap \not= b$}\cr
& &\R{ab}{ba}=r-r^{-1} , ~~~~~~~\mbox{\footnotesize
$a>b, \ap \not= b $}\cr
& &\R{a\ap}{\ap a}=(r-r^{-1})(1-\epsilon r^{\rho_a-\rho_{\ap}}) ,
{}~~~~~~\mbox{\footnotesize
$a>\ap $}\cr
& &\R{a\ap}{b \bp}=-(r-r^{-1})\epsilon_a\epsilon_b
r^{\rho_a-\rho_b} , ~~~~~~~
\mbox{\footnotesize $~~a>b ,~ \ap \not= b $}
\label{Rnonzero}
\ena
where $\epsilon=\epsilon_a \epsilon_{\ap}$, i.e.
 $\epsilon=1$ for $B_n$, $D_n$ and $\epsilon=-1$
for $C_n$.
\sk
{\sl Remark 1:}  The matrix $R$ has the following symmetry:
\eq
\R{ab}{cd}=\R{\cp\dpr}{\ap\bp}  \label{Rprimi}
\en

\indent {\sl Remark 2:}  If we denote by
$q,r$ the set of parameters $q_{ab},r$, we
have
\eq
R^{-1}_{q,r}=R_{q^{-1},r^{-1}} \label{Rprop1}
\en
\noi The inverse $R^{-1}$ is defined by
$\Rinv{ab}{cd} \R{cd}{ef}=\de^a_e \de^b_f=\R{ab}{cd}
\Rinv{cd}{ef}$.
Eq. (\ref{Rprop1}) implies
that for $|q|=|r|=1$, ${\bar R}=R^{-1}$.
\sk
{\sl Remark 3:} Let $R_r$ be the uniparametric $R$ matrix
for the $B, C, D$ q-groups. The multiparametric $R_{q,r}$
matrix is obtained from $R_r$ via the transformation
\cite{multiparam1,Schirrmacher}
\eq
R_{q,r}=F^{-1}R_rF^{-1}
\en
where $(F^{-1})^{ab}_{~~cd}$ is a diagonal matrix
in the index couples $ab$, $cd$:
\eq
F^{-1}\equiv diag (\sqrt{{r \over q_{11}}} ,
\sqrt{{r \over q_{12}}} , ... ~ \sqrt{{r \over q_{NN}}})
\label{effe}
\en
\noi where $ab$, $cd$ are ordered as
in the $R$ matrix.
Since $\sqrt{{r \over q_{ab}}} =(\sqrt{{ q_{ba}\over r}})^{-1}$
and $q_{a\ap}=q_{b\bp}$, the non diagonal
elements of $R_{q,r}$ coincide with those of $R_r$.
The matrix $F$ satisfies $F_{12}F_{21}=1$ i.e.
$F^{ab}{}_{ef}F^{fe}{}_{dc}=\delta^a_c\delta^b_d $,
the quantum Yang-Baxter equation $F_{12}F_{13}F_{23}
=F_{23}F_{13}F_{12}$ and the relations
$(R_r)_{12}F_{13}F_{23}=F_{23}F_{13}(R_r)_{12}$.
\sk
{\sl Remark 4:} Let $\Rh$ the matrix defined by
$\Rhat{ab}{cd} \equiv \R{ba}{cd}$.  Then the multiparametric
$\Rh_{q,r}$ is
obtained from $\Rh_r$ via the similarity transformation
\eq
\Rh_{q,r}=F\Rh_rF^{-1}
\en
The characteristic equation and the projector decomposition
of $\Rh_{q,r}$ are therefore the same as in the uniparametric case,
and we  have
\eq
(\Rh-rI)(\Rh+r^{-1}I)(\Rh-\epsilon r^{\epsilon-N} I)=0 \label{cubic}
\en
\eq
\Rh=r P_S - r^{-1} P_A+\epsilon r^{\epsilon-N}P_0  \label{RprojBCD}
\en
with
\eq
\begin{array}{ll}
&P_S={1 \over {r+\rminus}} [\Rh+\rminus I-(\rminus+\epsilon
r^{\epsilon-N})P_0]\\
&P_A={1 \over {r+\rminus}} [-\Rh+rI-(r-\epsilon r^{\epsilon-N})P_0]\\
&P_0= Q_N(r) K\\
&Q_N(r) \equiv (C_{ab} C^{ab})^{-1}={{1-r^2} \over
{(1-\epsilon r^{N+1-\epsilon})(1+\epsilon r^{-N+1+\epsilon})}}~,~~~~
K^{ab}_{~~cd}=C^{ab} C_{cd}\\
&I=P_S+P_A+P_0
\end{array}
\label{projBCD}
\en
Orthogonality  (and symplecticity) conditions can be
imposed on the elements $\T{a}{b}$, consistently
with  the $RTT$ relations (\ref{RTT}):
\eqa
& &C^{bc} \T{a}{b}  \T{d}{c}= C^{ad} I \nonumber\\
& &C_{ac} \T{a}{b}  \T{c}{d}=C_{bd} I \label{Torthogonality}
\ena
\noi where the (antidiagonal) metric is :
\eq
C_{ab}=\epsilon_a r^{-\rho_a} \de_{a\bp} \label{metric}
\en
\noi and its inverse $C^{ab}$
satisfies $C^{ab} C_{bc}=\de^a_c=C_{cb} C^{ba}$.
We see
that for the orthogonal series, the matrix elements of the metric
and the inverse metric coincide,
while for the symplectic series there is a change of sign:
$C^{ab}=\epsilon C_{ab}$. Notice also the symmetry
$C_{ab}=C_{\bp\ap}$.

The consistency of (\ref{Torthogonality}) with the $RTT$ relations
is due to the identities:
\eq
C_{ab} \Rhat{bc}{de} = \Rhatinv{cf}{ad} C_{fe} \label{crc1}
\en
\eq
 \Rhat{bc}{de} C^{ea}=C^{bf} \Rhatinv{ca}{fd} \label{crc2}
\en
\noi These identities
 hold also for $\Rh \rightarrow \Rh^{-1}$ and can be proved using
the explicit expression (\ref{Rnonzero}) of $R$.
\sk

We note the useful relations, easily deduced from (\ref{RprojBCD}):
\eq
C_{ab}\Rhat{ab}{cd}=\epsilon r^{\epsilon-N}C_{cd} ,~~~
C^{cd}\Rhat{ab}{cd}=\epsilon r^{\epsilon-N}C^{ab}  \label{CR}
\en
The metric $C$ can be used to express the symmetry property
(\ref{Rprimi})
in the covariant notation:
\eq
\R{ab}{cd}=C_{cp}C_{dq}\R{pq}{ef}C^{ea}C^{fb} =
C^{ae}C^{bf}\R{pq}{ef}C_{pc}C_{qd}~.
\en

The co-structures of the $B,C,D$ multiparametric quantum
groups have the same form as in the uniparametric case:
the coproduct
$\D$, the counit $\epsi$ and the coinverse $\kappa$ are given by
\eqa
& & \D(\T{a}{b})=\T{a}{b} \otimes \T{b}{c}  \label{cos1} \\
& & \epsi (\T{a}{b})=\delta^a_b\\
& & \kappa(\T{a}{b})=C^{ac} \T{d}{c} C_{db}
\label{cos2}
\ena
A conjugation (i.e. algebra antihomomorphism, coalgebra homomorphism
and involution, satisfying $\kappa(\kappa(T^*)^*)=T$)
can be defined
\sk
$\bullet$~~trivially as $T^*=T$. Compatibility with the
$RTT$ relations (\ref{RTT}) requires ${\bar R}_{q,r}=R^{-1}_{q,r}=
R_{q^{-1},r^{-1}}$,
i.e. $|q|=|r|=1$. Then the $CTT$ relations are invariant under
$*$-conjugation.  The corresponding real forms are
$SO_{q,r}(n,n;\Rbo)$, $SO_{q,r}(n,n+1;\Rbo)$
(for N even and odd respectively) and $Sp_{q,r}(n;\Rbo)$.
\sk
$\bullet$~~ via the metric
as
$T^*=(\kappa(T))^t$.
 The condition on $R$ is
${\bar \R{ab}{cd}}=\R{dc}{ba}$, which
happens for $q_{ab} {\bar q}_{ab}=r^2, r \in $ {\bf R}.
Again the $CTT$ relations are $*$-invariant.
The metric on a ``real" basis has compact signature
$(+,+,...+)$ so that the real form is  $SO_{q,r}(N;\Rbo)$.
\sk
$\bullet$~~as $(\T{a}{b})^*=\T{\ap}{\bp}$. This conjugation,
as far as we know, has never been discussed in the literature.
The conditions on $R$ are ${\bar \R{ab}{cd}}=\R{\bp\ap}{\dpr\cp}$,
and
due to (\ref{Rprimi})  they turn out to be
the same as for the preceding conjugation.
The compatibility with the $CTT$ relations follows from
$\bar C_{ab}=C_{\bp\ap}$ (when $r \in \Rb$).
\sk
$\bullet$~~
there is also a fourth way \cite{Cas2}
 to define a conjugation on the orthogonal
quantum groups $SO_{q,r}(2n,{\bf C})$, which extends to the
multiparametric case the one
proposed by
the authors of ref. \cite{Firenze1} for $SO_{q}(2n,{\bf C})$.
The conjugation is defined by:
\eq
(\T{a}{b})^*={\cal D}^a_{~c} \T{c}{d}
{\cal D}^d_{~b} \label{Tconjugation}
\en
\noi ${\cal D}$ being the matrix that
exchanges the index $n$ with the index $n+1$.
This conjugation is compatible with the coproduct: $\D (T^*)=(\D
T)^*$;
for $|r|=1$ it is also  compatible with the
orthogonality relations (\ref{Torthogonality}) (due to
${\bar C}=C^T$ and also $\Dc C \Dc = C$) and with the antipode:
$\kappa(\kappa(T^*)^*)=T$.  Compatibility with the $RTT$ relations
is easily seen to require
\eq
({\bar R})_{n \leftrightarrow n+1}=R^{-1}, \label{Rprop2}
\en
\noi which implies

i) $|q_{ab}|=|r|=1$
for $a$ and $b$ both different from $n$ or $n+1$;

ii) $q_{ab}/r \in {\bf R}$
when at least one of the indices $a,b$ is equal
to $n$ or $n+1$.
\sk
\noi This last conjugation leads to the
real form $SO_{q,r}(n+1,n-1;\Rbo)$, and
is in fact  the one  needed  to obtain $ISO_{q,r}(3,1;\Rbo)$, as
 discussed  in ref. \cite{Cas2} and later in this Letter.

\sk
Finally, we consider the $R$ matrix for the
$SO_{q,r}(N+2)$ and $Sp_{q,r}(N+2)$ quantum groups.
Using formula (\ref{Rmp}) or (\ref{Rnonzero}),
we find that it can be decomposed
in terms of  $SO_{q,r}(N)$ and $Sp_{q,r}(N)$ quantities
as follows (splitting the index {\small A} as
{\small A}=$(\circ, a, \bullet)$, with $a=1,...N$):
\eq
\R{AB}{CD}=\left(  \begin{array}{cccccccccc}
   {}&\circ\circ&\circ\bullet&\bullet
          \circ&\bullet\bullet&\circ d&\bullet d
      &c \circ&c\bullet&cd\\
   \circ\circ&r&0&0&0&0&0&0&0&0\\
   \circ\bullet&0&r^{-1}&0&0&0&0&0&0&0\\
   \bullet\circ&0&f(r)&r^{-1}&0&0&0&0&0&-\epsilon C_{cd} \lambda
r^{-\rho}\\
\bullet\bullet&0&0&0&r&0&0&0&0&0\\
\circ b&0&0&0&0&{r\over q_{\circ b}} \de^b_d&0&0&0&0\\
\bullet b&0&0&0&0&0&{r\over q_{\bullet b} }
\de^b_d&0&\lambda\de^b_c&0\\
a\circ&0&0&0&0&\lambda\de^a_d&0&{r \over q_{a \circ} } \de^a_c&0&0\\
a\bullet&0&0&0&0&0&0&0&{r\over q_{a \bullet}} \de^a_c&0\\
ab&0&-C^{ba} \lambda r^{-\rho}
&0&0&0&0&0&0&\R{ab}{cd}\\
\end{array} \right) \label{Rbig}
\en
\noi where $\R{ab}{cd}$ is the $R$ matrix for  $SO_{q,r}(N)$
or $Sp_{q,r}(N)$, $C_{ab}$ is the corresponding
metric,  $\lambda \equiv r-r^{-1}$,
$\rho={{N+1-\epsilon}\over 2}~(r^{\rho}=C_{\bullet \circ})$
and $f(r) \equiv \lambda (1-\epsilon r^{-2\rho})$.
The sign $\epsilon$ has been defined after eq. s
(\ref{Rnonzero}).
\sk

{\sl Theorem 1:} the quantum inhomogeneous groups
$\ISOqrN$ and $\ISpqrN$ are freely generated by the
non-commuting elements
\eq
{}~~\T{a}{b} , x^a , v , u\equiv v^{-1}
\mbox{ and the identity }  I~~~~~(a=1,...N)
\en
modulo the relations:

\eqa
& &\R{ab}{ef} \T{e}{c} \T{f}{d} = \T{b}{f} \T{a}{e} \R{ef}{cd}
\label{PRTT11}\\
& &\T{a}{b} C^{bc} \T{d}{c}=C^{ad} I \label{PRTT31}\\
& &\T{a}{b} C_{ac} \T{c}{d} = C_{bd} I \label{PRTT32}
\ena
\eqa
& &\T{b}{d} x^a={r \over q_{d\bullet}} \R{ab}{ef} x^e \T{f}{d}\\
& &\PA{ab}{cd} x^c x^d=0 \label{PRTT13}\\
& &\T{b}{d} v={q_{b\bullet}\over q_{d\bullet}} v \T{b}{d}\\
& &x^b v=q_{b \bullet} v x^b \label{PRTT15}\\
& &uv=vu=I \label{PRTT21}\\
& &u x^b=q_{b\bullet} x^bu\\
& &u \T{b}{d}={q_{b\bullet}\over q_{d\bullet}} \T{b}{d} u
\label{PRTT24}
\ena
\noi where $q_{a\bullet}$ are $N$ free complex parameters.
The matrix $P_A$ in eq. (\ref{PRTT13}) is the $q$-antisymmetrizer for
the
$B,C,D$ $q$-groups given by (cf.  (\ref{projBCD}):
\eq
\PA{ab}{cd}=- {1 \over {r+\rminus}}
(\Rhat{ab}{cd}-r\de^a_c \de^b_d + {r-r^{-1} \over
\epsilon r^{N-1-\epsilon} +1} C^{ab} C_{cd}) \label{PA}
\en
The co-structures are given by :
\eqa
& &\D (\T{a}{b})=\T{a}{c} \otimes \T{c}{b}  \label{Pcoproduct1}\\
& &\D (x^a)=\T{a}{c} \otimes x^c + x^a \otimes v\\
%& &\D (y_b)=y_c \otimes \T{c}{b}+u\otimes y_b \\
& &\D (v)=v \otimes v\\
& &\D (u)=u \otimes u
\ena
\eqa
& &\kappa (\T{a}{b})
=C^{ac} \T{d}{c} C_{db}=\epsilon_a\epsilon_b
r^{-\rho_a+\rho_{b}}~ \T{\bp}{\ap}\\
& &\kappa (x^a)
=- \kappa (\T{a}{c}) x^c u\\
& &\kappa (v)=  \epsilon u\\
& &\kappa (u)=\epsilon v
\ena
\eq
\epsi (\T{a}{b})=\de^a_b ~;~~\epsi (x^a)=0 ~;~~
\epsi (u)=\epsi (v)=\epsi(I)=1 \label{cfin}
\en
\sk
In the commutative limit $q\rightarrow 1 , r\rightarrow 1$ we
recover the algebra of functions
on $ISO(N)$ and $ISp(N)$ (plus the dilatation $v$ that can be set to
the
identity).

\sk
{\sl Proof :} our strategy will be to prove
 that the quantum groups $\ISOqrN$ and $\ISpqrN$
can be derived as the quotients
\eq
\frac{\SOqrNt}{H}~, ~~\frac{\SpqrNt}{H}
\label{quotient}
\en
\noi where $H$ is a suitable Hopf ideal in
 $\SOqrNt$ or $\SpqrNt$. Then the Hopf
structure of the groups in the numerators of (\ref{quotient})
is naturally inherited by the quotient groups \cite{Sweedler}.
We indicate by $\T{A}{B}$ the basic elements of
 $\SOqrNt$ or $\SpqrNt$, with the index convention
 {\small A}=$(\circ, a, \bullet)$, $a=1,...N$,
induced by the $R$ matrix of (\ref{Rbig}).
\sk
The space $H$ is defined
as the space of
all sums of monomials containing at least an element of the kind
$\T{a}{\circ}, \T{\bullet}{b}, \T{\bullet}{\circ}$.

We introduce the following convenient notations:
$\Tc$ stands for
$\T{a}{\circ}$,  $\T{\bullet}{b}$ or  $\T{\bullet}{\circ}$,
$\SqrNt$ stands for either $\SOqrNt$ or
$\SpqrNt$,
and we indicate by
$\Dtwo$, $\epsitwo$ and $\kappatwo$ the corresponding
co-structures.
\sk
We start the proof of {\sl Theorem 1} by proving
 first the important Lemma:
\sk
{\sl Lemma:} the space $H$ is a Hopf ideal in
 $\SqrNt$, that is, if

i) $H$ is a two-sided ideal in $\SqrNt$,

ii) $H$ is a co-ideal, i.e.
\eq
\Dtwo (H) \subseteq H \otimes \SqrNt + \SqrNt \otimes H;~~\epsitwo
(H)=0
\label{coideal}
\en

iii) $H$ is compatible with $\kappatwo$:
\eq
\kappatwo (H)\subseteq H~.\label{Hideal}
\en

\noi {\sl Proof:}
\sk
\noi ${}~~{}$i) $H$ is trivially a subalgebra of $\SqrNt$.
It is a right and left ideal since $\forall h\in H,\/ \forall a\in
\SqrNt ,~ha\in H \mbox{ and }
 ah\in H.$
This follows immediately from the definition of $H$ as sums of
monomials containing at least a factor $\Tc$.
$H$ is the
ideal in $\SqrNt$ generated by the elements $\Tc$.
\sk
\noi ${}~{}$ii) First notice that
$\Dtwo(\Tc)\in H \otimes \SqrNt + \SqrNt \otimes H .$  Now by
definition
of  $H$ we have
\eq
\forall ~h\in H ,~~~h=b\Tc c,~~~~~b,c \in \SqrNt .
\en
where $b\Tc c$ represents a sum of monomials.
Then we find
\eq
\Dtwo(h)=\Dtwo(b)\Dtwo(\Tc)\Dtwo(c) \in  H \otimes \SqrNt + \SqrNt
\otimes H ~.
\en
Moreover, since $\epsitwo$ vanishes on $\Tc$ we
have:
\eq
\epsitwo (h)= 0,~~~~\forall h \in H.
\en
These relations ensure that (\ref{coideal}) holds.
\sk

\noi ${}{}$iii)
\eqa
& &\kappatwo (\T{a}{\circ}) = C^{a\ap} \T{\bullet}{\ap} C_{\bullet
\circ}\\
& &\kappatwo (\T{\bullet}{b})=C^{\bullet \circ} \T{\bp}{\circ} C_{\bp
b}\\
& &\kappatwo (\T{\bullet}{\circ})=C^{\bullet\circ} \T{\bullet}{\circ}
C_{\bullet\circ}
\ena
\noi so that $\kappatwo (\Tc) \propto \Tc$ and therefore
\eq
\kappatwo (h)=\kappatwo(b\Tc
c)=\kappatwo(c)\kappatwo(\Tc)\kappatwo(b)
\in H
\en
and the {\sl Lemma} is proved. \square
\sk
Consider now the quotient
\eq
\frac{\SqrNt}{H}~,
\en
and the canonical projection
\eq
P ~:~~  \SqrNt\longrightarrow \SqrNt/{H}
\en
Any element of ${\SqrNt/H}$ is of the form $P(a)$. Also, $P(H)=0$,
i.e. $H=Ker(P)$.
\sk
Since $H$ is a two-sided ideal, ${\SqrNt/H}$ is an algebra with
the following sum and products:
\eq
P(a)+P(b)\equiv P(a+b) ~;~~ P(a)P(b)\equiv P(ab) ~;~~
\mu P(a)\equiv P(\mu a),~~~\mu \in \mbox{\bf C} \label{isoalgebra}
\en
We will use the following notation:
\eqa
&
&u\equiv P(\T{\circ}{\circ}),~~v\equiv
P(\T{\bullet}{\bullet}),~~z\equiv
P(\T{\circ}
{\bullet})\\
& &x^a \equiv P(\T{a}{\bullet}),~~y_a \equiv P(\T{\circ}{a})
\ena
and with abuse of symbols:
\eq
T^a{}_b\equiv P(T^a{}_b) ~;~~ I\equiv P(I) ~;~~ 0\equiv P(0)
\en
Using (\ref{isoalgebra}) it is easy to show that
$\T{a}{b}$, $x^a$, $y_b$, $u$, $v$, $z$
 and $I$
generate the algebra ${\SqrNt/H}$. Moreover from
the $RTT$  relations (\ref{RTT})
$R_{12}T_1T_2=T_2T_1R_{12}$ and the $CTT=C$
relations (\ref{Torthogonality})
in $\SqrNt$  we find the ``$P(RTT)$" and ``$P(CTT)$"
relations in ${\SqrNt/H}$:
\eq
P(R_{12}T_1T_2)=P(T_2T_1R_{12}) ~~~ i.e. ~~~
R_{12}P(T_1)P(T_2)=P(T_2)P(T_1)R_{12}  \label{PRTT}
\en
\eq
P(CTT)=C~~~i.e.~~~CP(T)P(T)=C \label{PCTT}
\en

{\sl Proposition :} The projected relations (\ref{PRTT}) and
(\ref{PCTT}) are equivalent to  the relations
(\ref{PRTT11})-(\ref{PRTT24}),
supplemented by the two constraints:
\eq
y_b=-r^{\rho} \T{a}{b} C_{ac} x^c u \label{ipsilon}
\en
\eq
z=-{1\over {(\epsilon r^{-\rho}+r^{\rho-2})}} x^b C_{ba} x^a u
\label{PRTT44}
\en

{\sl Proof : }  Consider the $R$ matrix decomposition (\ref {Rbig}) .
The three
kinds of indices $\circ , a ,
\bullet$  yield  81  $RTT$ relations.
Out of these only 41 are independent and give all the
$q$-commutations between the $T^A{}_B$ elements:
they contain all the information of the $RTT$ relations.
We then project these 41 relations to obtain the $P(RTT)$ relations.
We proceed in a similar way with the 9
$CTT$ relations to obtain the $P(CTT)$ relations:
in particular one finds $uv=vu=I$.

The projected relations obtained in this way are not independent.
In fact  choosing  the lower indices in the
 $P(CTT)$ relations as  $(\bullet d)$  we find
the constraint (\ref{ipsilon}).
All the projected  relations that contain the elements $y$ are
a consequence of the  remaining projected relations
and of (\ref{ipsilon}).
Therefore (\ref{ipsilon}) is a consistent constraint.
The contraction of the $({}^a{}_{\bullet} {}^b{}_{\bullet})$ $P(RTT)$
relations with the metric $C_{ab}$ gives (\ref{PRTT44})
[use (\ref{CR}) and  $C_{ab}C^{ab}$ in (\ref{projBCD})] .
All the other projected relations  containing the element $z$
are a consequence of
the remaining ones  and of (\ref{PRTT44}).
Finally all the
projected relations containing the element $u$ are a
consequence of $u=v^{-1}.$

We thus arrive at the minimal set of $P(RTT)$ and $P(CTT)$
relations given by  (\ref{PRTT11})-(\ref{PRTT24}),
(\ref{ipsilon}) and (\ref{PRTT44}).
The  {\sl Proposition} is then proved. \square
\sk

This implies that we can choose as independent generators
the set $\T{a}{b}$, $x^a$, $v$, $u\equiv v^{-1}$, and $I$.
\sk
Let us indicate by $\Dtwo, \epsitwo$ and $\kappatwo$ the costructures
of
$S_{q,r}(N+2)$, defined by:
\eqa
& &\Dtwo(\T{A}{B})=\T{A}{C} \otimes \T{C}{B}\\
& &\kappatwo (\T{A}{B})=C^{AC} \T{D}{C} C_{DB} \\
& &\epsitwo (\T{A}{B})=\de^A_B \label{costructuresbigtwo}
\ena

Since H is a Hopf ideal then ${\SqrNt/H}$ is also a Hopf
algebra with co-structures:
\eq
\D (P(a))\equiv (P\otimes P)\Dtwo(a) ~;~~ \epsi(P(a))
\equiv\epsitwo(a) ~;~~
\kappa(P(a))\equiv P(\kappatwo(a)) \label{co-iso}
\en
Indeed (\ref{coideal}) and (\ref{Hideal}) ensure
that $\D ,~\epsi ,$
and $\kappa$ are well defined. For example
\eq
(P\otimes P)\Dtwo(a) = (P\otimes P)\Dtwo(b)~~
\mbox{ if }~~ P(a)=P(b) ~. \label{welldefined}
\en
In order to prove the Hopf algebra axioms of  the Appendix  for
$\D,~\epsi,~\kappa$ we just have to project
those for $\Dtwo,~\epsitwo,~\kappatwo~.$
For example, the first axiom is proved by applying
$P\otimes P \otimes P$ to $(\Dtwo \otimes id)\Dtwo(a) =
 (id \otimes \Dtwo)\Dtwo(a)$. The other axioms are proved
in a similar way.
\sk
In conclusion, the elements
$\T{a}{b}$, $x^a$, $v$, $u\equiv v^{-1}$ and $I$
generate the Hopf algebra ${\SqrNt/H}$ and satisfy the
$P(RTT)$ and $P(CTT)$ relations
(\ref{PRTT11})-(\ref{PRTT24}). The
co-structures defined in (\ref{co-iso}) act on them exactly as the
co-structures defined in
(\ref{Pcoproduct1})-(\ref{cfin}). Therefore the explicit structure of
the Hopf
algebra
${\SqrNt/H}$ is the one described in {\sl Theorem 1}.
We have
\eq
\ISqrN = \frac{\SqrNt}{H}  ~,
\en
and {\sl Theorem 1} is proved.  \square \square
\sk

The canonical projection $ P ~:~~ \SqrNt \rightarrow
\ISqrN$ is an
epimorphism between these two Hopf algebrae.

\sk

\noi {\sl Note 1:}  the consistency of the $P(RTT)$ and $P(CTT)$
relations with the co-structures $\D,\epsi$ and $\kappa$
is easily proved. For example,
\eq
\D (P(R_{12}T_1T_2)-P(T_2T_1R_{12}))=0
\en
\noi is a particular case of eq.  (\ref{welldefined}).
Similarly for
$\epsi$ and $\kappa$, and for the $P(CTT)$
relations.
\sk
We are now able to give a $R$ matrix formulation of the
 inhomogeneous $\ISOqrN$ and $\ISpqrN$
q-groups.
Indeed recall that $S_{q,r}(N+2)$ is the Hopf algebra freely
generated by the
non-commuting matrix elements
$T^A{}_B$ modulo the ideal  generated by the  $RTT$ and $CTT$
relations [$R$ matrix and metric $C$ of $S_{q,r}(N+2)$].
This can be expressed as:
\eq
S_{q,r}(N+2)\equiv \frac{<T^A{}_B>}{[RTT, CTT]}
\en
Therefore we have (recall that
$H\equiv[\T{a}{\circ},\T{\bullet}{b},\T{\bullet}{\circ}]
\equiv [\Tc]$) :
\eq
\ISqrN=  \frac{\SqrNt}{[\Tc]}
=\frac{ <T^A{}_B>/[RTT, CTT] }{[\Tc]}=
\frac{<T^A{}_B>}{[RTT, CTT, \Tc]}
\en
So that we have shown the following
\sk
{\sl Theorem 2:} the quantum inhomogeneous groups
$\ISOqrN$ and $\ISpqrN$ are freely generated by the
non-commuting matrix elements $\T{A}{B}$
 [{\small A}=$(\circ, a, \bullet)$, with $a=1,...N$)] and the
identity
$I$,
modulo the relations:
\eq
\T{a}{\circ}=\T{\bullet}{b}=\T{\bullet}{\circ}=0 , \label{Tprojected}
\en
\noi the $RTT$ relations
\eq
\R{AB}{EF} \T{E}{C} \T{F}{D} = \T{B}{F} \T{A}{E} \R{EF}{CD},
\label{RTTbig}
\en
\noi and the orthogonality (symplecticity) relations
\eqa
& &C^{BC} \T{A}{B}  \T{D}{C}= C^{AD} \nonumber\\
& &C_{AC} \T{A}{B}  \T{C}{D}=C_{BD} \label{CTTbig}
\ena
The co-structures of $\ISOqrN$ and $\ISpqrN$
are simply given by:
\eqa
& &\D (\T{A}{B})=\T{A}{C} \otimes \T{C}{B}\\
& &\kappa (\T{A}{B})=C^{AC} \T{D}{C} C_{DB} \\
& &\epsi (\T{A}{B})=\de^A_B \label{costructuresbig}
\ena

{\sl Note 2 :} the $\T{A}{B}$
matrix elements in eq. (\ref{RTTbig})
are really a  {\sl redundant} set:  indeed
not all of them are independent, see the constraints
(\ref{ipsilon}) and (\ref{PRTT44}).  This is necessary
if we want to express the $q$-commuations
of the $\ISOqrN$ and $\ISpqrN$ basic group elements
as $RTT=TTR$  (i.e. if we
want an $R$-matrix
formulation).  Remark that,
in the $R$-matrix formulation
for $\IGLqrN$, all  the $\T{A}{B}$
are independent \cite{Cas3,Aschieri1}.
\sk
{\sl Note 3 :} From the commutations
(\ref{PRTT11}) - (\ref{PRTT24})
 we see that
one can set $u=I$ only when $q_{a\bullet}=1$ for all $a$.
{}From $q_{a\bullet} = r^2 /q_{\ap\bullet}$, cf. eq. (\ref{qab2}),
this
implies also $r=1$, in agreement with the results of
ref. \cite{Cas2}, where a differential calculus for
$ISO_{q,r}(N)$ without dilatations was found only for
$r=1$.
\sk
{\sl Note 4} : eq.s (\ref{PRTT13}) are the multiparametric
(orthogonal
or symplectic) quantum plane commutations. They follow
from the   $({}^a{}_{\bullet} {}^b{}_{\bullet})$  $P(RTT)$ components
and (\ref{PRTT44}).
\sk
Finally, it is not difficult to see how the real forms of $\SqrNt$
are
inherited by $\ISqrN$. In fact, only the first
and the fourth real forms of $\SqrNt$, discussed after (\ref{cos2}),
are
compatible with the coset structure of
$\ISqrN$. More precisely, $H$ is a $*$-Hopf ideal, i.e.
$(H)^* \subseteq H$, only for
$T^*=T$ or $(\T{a}{b})^*={\cal D}^a_{~c} \T{c}{d}
{\cal D}^d_{~b}$. Then we can define a $*$-structure on $\ISqrN$ as
$[P(a)]^* \equiv P(a^*),~\forall a \in \SqrNt$.
The conditions on the parameters are respectively:
\sk

$\bullet$~~ $|q_{ab}|=|q_{a\bullet}|=|r|=1$ for $ISO_{q,r}(n,n;\Rb)$,
$ISO_{q,r}(n,n+1;\Rb)$
and $ISp_{q,r}(n;\Rb)$.
\sk
$\bullet$~~For $ISO_{q,r}(n+1,n-1;\Rb)$ : $|r|=1$; $|q_{ab}|=1$ for
$a$
and $b$
both different from $n$ or $n+1$;  $q_{ab}/r \in {\bf R}$
when at least one of the indices $a,b$ is equal
to $n$ or $n+1$;   $q_{a\bullet}/r \in \Rb$ for $a=n$ or $a=n+1$.
\sk
In particular,  the quantum Poincar\'e group $ISO_{q,r}(3,1;\Rb)$
is obtained by setting $|q_{1\bullet}|=|q_{2\bullet}|=|r|=1$,
$q_{12} /r \in \Rb$.
\sk\sk

%%%%%%%%%%%%%%%%%%%%%%%%%%%%%%%%%%%%%%%%%

\centerline{\bf APPENDIX :  the Hopf algebra axioms}

%%%%%%%%%%%%%%%%%%%%%%%%%%%%%%%%%%%%%%%%%
\sk
A Hopf algebra over the field $K$ is a
unital algebra over $K$ endowed
with the  linear maps:
\eqa
& &\D~: ~~A \rightarrow A\otimes A \\
& &\epsi~:~~ A\rightarrow K\\
& &\kappa~:~~A\rightarrow A
\ena
\noi satisfying the following
properties $\forall a,b \in A$:
\eq
(\D \otimes id)\D(a) = (id \otimes \D)\D(a)
\en
\eq
(\epsi \otimes id)\D(a)=(id \otimes \epsi)\D(a) =a \label{axiom2}
\en
\eq
m(\kappa\otimes id)\D(a)=m(id \otimes \kappa)\D(a)
=\epsi(a)I \label{kappadelta}
\en
\eq
\D(ab)=\D(a)\D(b)~;~~\D(I)=I\otimes I
\en
\eq
\epsi(ab)=\epsi(a)\epsi(b)~;~~\epsi(I)=1
\en
\noi where $m$ is the multiplication map $m(a\otimes b)
= ab$.
{} From  these axioms we deduce:
\eq
\kappa(ab)=\kappa(b)\kappa(a)~;~~\D[\kappa(a)]=\tau(\kappa\otimes
\kappa)\D(a)~;~~\epsi[\kappa(a)]=\epsi(a)~;~~\kappa(I)=I
\en
where $\tau(a \otimes b)=b\otimes a$ is the twist map.
\sk\sk\sk

\vfill\eject
\end{document}